\documentclass[10pt,,notitlepage,superscriptaddress,preprintnumbers,amsmath,amssymb,nofootinbib,aps,prd]{revtex4-1}
\usepackage{amsmath,amssymb,amsfonts}
\usepackage{bm}
\usepackage{graphicx,color}
\usepackage{multirow}
\usepackage{dcolumn}
\usepackage{hyperref}
\usepackage{appendix}
\usepackage{ulem}
\begin{document}
	
	\title{A spherically symmetric gravitational solution of nearly conformally flat metric~measure~space}
	\author{S.Oghbaiee}
	\email{s_oghbaiee@sbu.ac.ir}
	\affiliation{Department of Physics, Shahid Beheshti University, Tehran 1983969411, Iran}
	\author{N. Rahmanpour}
	\email{nrahmanp@usc.edu}
	\affiliation{Department of Computer Science, University of Southern California Los Angeles, CA, USA}
	\author{H. Shojaie}
	\email{h-shojaie@sbu.ac.ir}
	\affiliation{Department of Physics, Shahid Beheshti University, Tehran 1983969411, Iran}
	\begin{abstract}
	In this manuscript, we study the nearly flat approximation of a conformally invariant gravitational theory in metric measure space (MMS). In addition, we investigate the transformation of the energy-momentum tensor in this context  and obtain the vacuum solution of MMS and  its weak field limit in the spherically symmetric coordinates. We show that while it is already a vacuum solution, it can simulate dark matter when restricted to the framework of general relativity, i.e., a  symmetry-broken conformal frame. This is done by employing a density function which is an essential part of MMS. We derive an equation for the density function for a general profile of a rotation curve obtained from observations. Specifically, the density function corresponding to two well-known profiles PSS and NFW are provided.
	\begin{description}
   \item[Keywords]
    Metric measure space; Conformal gravity; Linearized gravity; Flat rotation curve.
  \end{description}
	\end{abstract}
	\maketitle
	
\section{Introduction}
	
The elegant simplicity in formulation, following a delicate arrangement of deeply thoughtful principles, together with a robust mathematical framework, all backed by an ingenious way of reasoning, has made Einstein's theory of gravitation, namely general relativity (GR), a relatively special theory of all time. Although on the way of its construction, the author's aim was mainly focused on the generalization of the concept of the observers rather than justifying little observational evidence available theretofore~\cite{einstein1915preussische}, surprisingly, it has so far outdistanced all of its old and new competitors not only in successfully passing obtainable tests from weak to strong regimes but in intriguingly making predictions of gravitational phenomena unknown at the time~\cite{will2014confrontation}.
	
Among all features of GR, observer independence is maybe the most important aspect of the theory. This generalizes the concept of inertial observers to all physical observers and even beyond, to formulate underlying physics in any arbitrary and bias-free framework. This property of the theory, dubbed diffeomorphism invariance, seems to be a generic property of every geometry-based theory of physics. However, we are not sure whether the roads to formulate theories in this way would be paved clear ones. This may be the origin of unsatisfactory inconsistencies between GR and quantum mechanics (QM); inconsistencies with models formulated in the latter, such as elementary particle physics, as well as recipes in this context like Quantum Field Theories (QFT).
	
On the other hand, regardless of these inconsistencies, some hints motivate one to step forward and think of possible improvements to GR. It is not very hard to mention some deficiencies in physics that GR may be accountable for. From the observational point of view, for instance, the justification of dark energy, a clear explanation for dark matter, and a proposal to overcome $H_0$ tension are some topics in this context. Although GR is not credited for all these problems, it should not be plausibly regarded as entirely irresponsible; GR has a few exact solutions, and almost always, when applied to the real world, the approximations come to the stage.
	
Regardless of this seemingly impenetrable barrier towards unification, GR has never been historically regarded as an ultimate version of a theory of gravitation, even not by its inventor. This can be witnessed by first attempts mainly focused on the constructions of more general geometrized theories to include Electrodynamics (EM) as well~\cite{weyl1918gravitation,weyl1929electron}. The non-stop efforts towards unification have had many ups and downs, led to the rising of many hypotheses and theories, and have always been in the mainstream of theoretical physics. The geometrization of other fundamental forces in the context of gauge theories and the emergence of theories and contexts such as String Theory and Loop Quantum Gravity (LQG) are just a few examples of these endeavors.
	
All these lead to the question of whether GR is a special and symmetry-broken case of a generalized theory with a larger symmetry group. It is worth noting that in physics, all equations have global symmetry with respect to unit invariance, regardless of whether they are dimensionful or not\footnote{The unit change, when physical quantities, parameters, and constants are considered, is better to be called unit covariance instead of unit invariance, since these things are not necessarily invariant, but they change covariantly; i.e., consistently.}. However, like many other global symmetries, this global symmetry does not generally have physical significance. Nevertheless, in theories with local unit invariance, where by local units one means units that are functions of spacetime coordinates, the physical consequences \textit{may} emerge. Attempts to construct such theories date back to the introduction of the Weyl geometry, later revised by Dirac~\cite{weyl1919neue,dirac1973long}. In particular, in theories with a dimensionful coupling constant, this symmetry can be interpreted as gauge invariance, namely scale invariance. This is because this invariance is responsible for an internal symmetry rather than an external one. Then, fixing the units, being a kind of gauge fixing, eliminates the redundant degrees of freedom and leaves only the physical ones. The price one pays is that one encounters a broken symmetry. However, it is not unusual throughout physics; for example, when the gravitational waves (GWs) are considered, full diffeomorphism invariance has not been preserved anymore.
	
Although the scale invariance and conformal invariance are not generally the same, a theory in the latter class will respect the former~\cite{polchinski1988scale,nakayama2015scale}. There are different ways to construct a theory that respects conformal invariance (see, for example,~\cite{faci2013constructing} and references therein). To generalize the aspect of the conformal invariance to curved spacetime, however, it seems that the most general symmetry that preserves this is the Weyl invariance in which metric transforms as $g_{\mu\nu} \rightarrow e^{2 w} g_{\mu\nu}$~\cite{farnsworth2017weyl}. But Weyl symmetry is not generally the symmetry of physics (especially of GR). To recover the conformal invariance and eliminate the effect of the Weyl transformation, a scalar field, usually called "Weyl compensator"~\cite{shaukat2010weyl}, is introduced like what is done in~\cite{gover2009weyl}. The joint transformation is traditionally called conformal transformation. When applied to GR, it extends the group of symmetry to $Weyl \rtimes Diff$.
	
Metric Measure Space (MMS), alternatively, as a generalization of (pseudo-)Riemannian manifolds, is one of such frameworks which lets one build up conformal invariant objects out of the Riemannian-invariant ones. It incorporates the Weyl transformation in addition to the transformation of a compensator scalar field, namely the density function, which relates a more general measure to the natural volume element by $dm=e^{-f} dvol(g)$. In~\cite{rahmanpour2015conformally,rahmanpour2016metric} following~\cite{chang2006conformal}, the authors provide a conformally invariant theory of gravitation which is basically consistent with GR, since when the density function $f$ is fixed to be a constant function, GR is retrieved. In another part of their work, they investigate similarities and differences between MMS and Weyl integrable space (WIS). For instance, they show that in contrast to WIS, the metricity and integrability are maintained in MMS. As a result, Rahmanpour \& Shojaie suggest MMS as an alternative to WIS  as a platform to construct conformally invariant theories of gravitation~\cite{rahmanpour2015conformally}. Indeed, despite the fact that these two mathematical frameworks, namely MMS and WIS, are constructed based on different assumptions, the authors have shown that some relations, such as the contracted second Bianchi identity, are entirely similar in these spaces. It is an interesting result from both mathematical and physical perspectives.
    
This manuscript investigates the other aspects of the conformally invariant theory of gravitation in MMS. To do so, first, in Sect~\ref{IntroMMS}, MMS is introduced in more detail and the structure of the theory is reviewed. Since a general tensor may depend on fields in addition to or rather than metric, recipes for constructing conformally invariant tensors may not be the same. However, a vital case in point in a gravitational theory is the energy-momentum tensor, which includes matter field as well. In Sect.~\ref{E-M tensor}, as an example of this type, we study the energy-momentum tensor of a perfect fluid in MMS.
In Sect.~\ref{Linearized theory}, we present a linearized version of the conformally invariant theory in the weak field limit.  Indeed, as gravity is a weak force we usually need to use the full nonlinear equations in the vicinity of massive objects like black holes. Perturbation methods, based on Taylor series expansion, mainly rely on the superposition of terms with different orders of magnitude to approximate a solution to a differential equation.  Linearization is a name for considering perturbations up to first order.  GR and its modifications are non-linear theories and do not have many exact solutions, highlighting the need for approximation techniques including perturbation methods. 

In Sect.~\ref{Spherical}, we obtain the spherically symmetric vacuum solution of the theory. We intend to use this solution to provide an explanation for the flat rotation curves (FRC) of galaxies in Sect~\ref{RotCur}. FRC problem is a discrepancy between the theoretical and the observational velocities of outer stars in galaxies~\cite{oort1932force,sofue2001rotation}. Indeed, there are different models in the literature trying to solve this problem~\cite{mannheim1989exact,mannheim2006alternatives,lobo2008dark,capozziello2006dark,mbelek2004modelling,mcgaugh2016radial,moffat2006scalar,rahvar2014observational}. We provide an alternative method to explain FRC in the context of MMS. To accomplish that, we need a corresponding density function to justify FRC. To do so, we employ the Persic--Salucci--Stel velocity profile~\cite{persic1996universal} and the Navarro--Frenk--White density profile~\cite{navarro1996structure}, to find a suitable density function. The last section, i.e., Sect~\ref{ConRem}, is devoted to some remarks. It is worth noting that despite using the results of~\cite{persic1996universal} and~\cite{navarro1996structure}, no datasets have been generated or analyzed in this manuscript.

One final point to remember is that the research has been carried out in the geometrized unit where the gravitational constant and the speed of light are set to be the unity ($G=c=1$).

\section{Metric Measure Space PRELIMINARIES}\label{IntroMMS}
	
On every $n$-dimensional (pseudo-)Riemannian manifold $(M^{n},g)$, the volume element $dvol(g)$ is defined naturally. To generalize the concept of volume, one can promote a (pseudo-)Riemannian manifold to a metric measure space (MMS). In this case, MMS is the triplet $(M^{n},g,m)$ with $(M^n,g)$ and $m$ being a (pseudo-)Riemannian manifold and the generalized smooth measure, respectively. The measure $m$ is related to the natural volume element by $dm=\exp(-f) dvol(g)$ where $f$ is a smooth function on $M^n$ called the density function.  Henceforth in this manuscript, as we are concerned with physical theories, we work with pseudo-Riemannian manifolds, and we use the name MMS instead of pseudo-Riemann measure space.
		
In 2005, in an article by Chang et al.~\cite{chang2006conformal}, it is shown that MMS is compatible with conformal transformations. In this space, conformal transformation is applied by a joint transformation
\begin{equation}\label{joint trans}
	\left\{
	\begin{array}{l}
		\hat{g}_{\mu \nu}= e^{2\omega} g_{\mu \nu} \\
		\\
		\hat{f}=f+(n-s') \omega,
	\end{array} \right.
\end{equation}
of metric $g$ and the density function $f$, respectively. In the latter, $s'$ is the conformal weight of measure $dm$. According to their method, one can construct a conformally invariant tensor of weight $ s$, shown by $\mathcal{I}_{s}(g,f)$, from its Riemannian invariant counterpart $I(g)$. Indeed,  $I(g)$ can be any geometrical tensor in the Riemannian space. It is natural to expect two essential features for the conformally invariant tensor $\mathcal{I}_{s}(g,f)$. First,  $\mathcal{I}_{s}(g,0)=I(g)$, and second, $\mathcal{I}_{s}(\hat{g},\hat{f})= \exp(s\omega) \mathcal{I}_{s}(g,f)$. Then it is easy to show that 
\begin{equation}\label{mod tensor}
	\mathcal{I}_{s}(g,f)=\exp \left(\frac{f s}{n-s^{\prime}}\right) I\left(\exp \left(\frac{-2 f}{n-s^{\prime}}\right) g\right),
\end{equation}
satisfies the desired features. This relation can be rewritten elegantly by introducing the ``canonical metric'' $g_m$. The canonical metric $g_m$ is defined as  $g_{m}=\exp(\frac{-2f}{n-s'})g$, and is obtained by setting $\hat{f}=0$ into the joint transformation. Accordingly, the Eqn.~(\ref{mod tensor}) can be demonstrated by
\begin{equation}\label{mod Ten canonic}
	\mathcal{I}_{s}(g,f)=\exp \left(\frac{f s}{n-s^{\prime}}\right) I(g_m) .
\end{equation}
	
It is worth noting that there is freedom to choose the conformal weight of the tensors, including the conformal weight of $dm$, namely $s'$. For instance, Chang et al. in~\cite{chang2006conformal}, set $s'=0$ for convenience, while Rahmanpour \& Shojaie in~\cite{rahmanpour2016metric} take $s'=2$ as they were interested to draw an analogy between MMS and WIS. Hereafter, we also set $s'=2$ and confirm that this will not affect the results. In addition, it should be mentioned that the conformal weights of some tensors are mutually dependent. For example, the conformal weights $s_0$ and $s_1$ of the modified counterparts of the Ricci tensor and the Ricci scalar, namely $\mathcal{R}_{\mu\nu}$ and $\mathcal{R}$ respectively, are related via $s_{1}=s_{0} - 2$. Therefore within four-dimensional MMS, this relation leads to
\begin{equation}\label{mod R ten}
	\begin{aligned}
		\mathcal{R}_{\mu \nu}(g,f) &=\exp \left(\frac{s_0 f}{2}\right)\left[R_{\mu \nu}(g)+ \nabla_{\mu} \nabla_{\nu} f+\frac{1}{2} \nabla_{\mu} f \nabla_{\nu} f\right.\\
			&\left.+\frac{1}{2} \nabla^{\rho} \nabla_{\rho} f g_{\mu \nu}-\frac{1}{2} \nabla^{\rho} f \nabla_{\rho} f g_{\mu \nu}\right],
	\end{aligned}
\end{equation}
\begin{equation}\label{mod R sca}
		\mathcal{R}(g,f)=\exp \left(\frac{\left(s_1 + 2\right) f}{2}\right)\left[R(g)+3 \nabla^{\mu} \nabla_{\mu} f-\frac{3}{2} \nabla^{\mu} f \nabla_{\mu} f\right].
\end{equation}
	
The theory introduced in~\cite{rahmanpour2015conformally,rahmanpour2016metric}, is a gravitational theory in which the conformal invariance, in addition to diffeomorphism invariance, holds both at the level of the action and at the level of the equation of motion. To have a conformally invariant weightless action in MMS
\begin{equation}\label{action}
		\int dm \left(\mathcal{R}(g,f)+ 8 \pi \mathcal{L}_{M}(\psi,g,f) \right) ,
\end{equation}
one must set $s_{1}=-s'$. Afterward, using the variational method and introducing a modified energy-momentum tensor $\mathcal{T}_{\mu \nu}$, they obtain Einstein's field equation counterpart in MMS as 
\begin{equation}\label{mod G}
		\mathcal{G}_{\mu \nu}(g,f) \equiv \mathcal{R}_{\mu \nu}(g,f) - \frac{1}{2} \mathcal{R}(g,f) g_{\mu \nu} = 8 \pi \mathcal{T}_{\mu \nu}(\psi , g , f) .
\end{equation}
	
Regarding Eqn.~(\ref{mod G}), it is worth mentioning that tensor fields, such as the energy-momentum tensor, which depend on fields in addition to the metric, do not generally transform via Eqn.~(\ref{mod tensor}) by a conformal transformation. In other words, the suitable transformation of a field is restrictively based on the other fields on which it depends  and the way the tensor is physically defined. This may lead to observable discrepancies between different conformal frames, as that observable may be a conformally frame-dependent observation. A more detailed discussion with an example is provided in Sect~\ref{E-M tensor}.

Another aspect of MMS, which is directly related to the notion of generalized measure, is that any quantity or relation which is defined through an integral over volume {\it may} change. In particular, although the covariant derivative in this space is the same as the covariant derivative in Riemannian geometry, it is not necessarily the case for the other differential operators. For example, it can be shown that the divergence operator in MMS is not simply the contraction of the covariant derivative with the tensor it acts upon\cite{rahmanpour2015conformally}. As a result, the invariance of the geometrical part of the action under the diffeomorphism leads to a generalized contracted second Bianchi identity. The invariance of the matter part of the action under the diffeomorphism on the other hand, yields a generalized identity for the conservation of energy-momentum tensor. Using this identity, the authors of~\cite{rahmanpour2015conformally} obtain a conformally invariant geodesic equation in the context of MMS.
	
In summary and in the four-dimensional Lorentzian manifold, the equations of motion~(\ref{mod G}) are invariant under transformations within the class
\begin{equation}\label{class of g}
		[(g_m,0)]=\{ (g,f) \lvert g = \exp(2\omega) g_m , f = 2\omega \} 
\end{equation}
as well as general coordinate transformations, where $g_m$ is the canonical metric with which GR is retrieved; i.e.
\begin{equation}\label{reduced G}
	 \exp(-\frac{s f}{2}) \mathcal{G}_s (g,f)= \mathcal{G}_s (g_m,0) =  G (g_m)
\end{equation}
with $G$ and $s$ being the traditional Einstein tensor and its conformal weight, respectively. In other words, each class of type~(\ref{class of g}) presents a solution of MMS. By substituting Eqns.~(\ref{mod R ten}) and~(\ref{mod R sca}) in the LHS of Eqn.~(\ref{mod G}), it is reduced to
\begin{equation}\label{separated G}
		\mathcal{G}_s (g,f)= \exp(\frac{s_0 f}{2}) \left(G_{\mu\nu}(g) - 8 \pi \mathcal{H}_{\mu\nu}(g,f)\right).
\end{equation}
From Eqns.~(\ref{reduced G}) and~(\ref{separated G}), it can easily be inferred that the conformal weight of the Einstein tensor is the same as that of the Ricci tensor, i.e., $s=s_0$. A simple comparison shows that
\begin{equation}\label{def H}
	\mathcal{H}_{\mu\nu}(g,f) \equiv -\frac{1}{8 \pi}  \left( \nabla_{\mu} \nabla_{\nu} f+\frac{1}{2} \nabla_{\mu} f \nabla_{\nu} f -\nabla^{\rho} \nabla_{\rho} f g_{\mu \nu}+\frac{1}{4} \nabla^{\rho} f \nabla_{\rho} f g_{\mu \nu} \right)
\end{equation}
which resembles the effect of some sort of mass distribution. It is evident that a constant $f$ corresponds to a global Weyl rescaling and is interpreted as the global change of unit. In this case, $\mathcal{H}=0$, and no physical consequence occurs. Thus, without loss of generality, we can set $f=0$ if it is constant.  A more detailed discussion on the energy-momentum tensor is provided in the following section.

\section{Energy--Momentum Tensor in metric measure space}\label{E-M tensor}

In the MMS theory of gravitation, there is a straightforward recipe to construct conformally invariant quantities out of Riemannian diffeomorphism invariant counterparts, i.e. Eqn.~(\ref{mod Ten canonic}). However, this instruction is restricted to tensors that solely depend on metric and its derivatives. Generally, contributions of other fields (in addition to metric) as sources of constructing a tensor field, demand a more careful treatment and probably different instructions to build up a conformally invariant tensor field. In other words, it is plausible that the most suitable instruction mainly depends on how a field is defined. In particular, the energy-momentum tensor is built upon measurements either in labs or via (cosmological) observations.  As an example of this tensor, consider the perfect fluid, which in addition to the metric, depends on the matter field via the matter density $\rho$ and the pressure $P$, in the form
\begin{equation}\label{E-M Ten}
    T_{\mu\nu} = \left( \rho + P \right) u_\mu u_\nu + P g_{\mu\nu} .
\end{equation}
It is clear that there is no explicit derivative of any kind in Eqn.~(\ref{E-M Ten}) that might cause the derivatives of the density function $f$ to emerge. More precisely, the Einstein equations in the canonical frame and in another arbitrary one, are related as
\begin{equation}\label{E-M in MMS}
\begin{aligned}
        8 \pi T (\psi_m , g_m) =  G(g_m) &= \mathcal{G}_s (g_m , 0) \\
        &= \exp(-\frac{s f}{2}) \mathcal{G}_s (g , f) = 8 \pi \exp(-\frac{s f}{2}) \mathcal{T}_s (\psi , g , f),
\end{aligned}
\end{equation}
with 
\begin{equation}
	\psi = \exp ( \frac{s_{\psi} f}{2} ) \psi_{m} ,
\end{equation}
where $\psi_m$, $\psi$, and $s_{\psi}$ are matter fields in the above conformal frames and the corresponding conformal weight, respectively~\cite{blagojevic2001gravitation}.

On the other hand, a straightforward calculation employing Eqns.~(\ref{reduced G}) and~(\ref{separated G}), for a perfect fluid, leads to  
\begin{equation}\label{E-M in Riem}
	8 \pi T (\psi_m , g_m) =  G(g_m) = G(g)- 8 \pi \mathcal{H}(g , f) = 8 \pi  T (\psi, g) .
\end{equation}
In fact, when a conformal transformation is applied, the change in metric and consequently in the proper velocity $u$ is compensated by the change in the units of the matter density and the pressure; any local changes in scale can be interpreted as local changes in the units. That is, the difference between $\psi_m$ and $\psi$ is in the different local units used to specify them. Comparing Eqns.~(\ref{E-M in MMS}) and~(\ref{E-M in Riem}), one can conclude
\begin{equation}\label{Tmms-T}
    \mathcal{T}_{s} (\psi, g ,  f) =  \exp(\frac{s f}{2}) T (\psi_m , g_m)= \exp(\frac{s f}{2}) T (\psi, g).
\end{equation}
However, recalling the lines after Eqn.~(\ref{mod Ten canonic}) for setting $s' = 2$ in compatibility with WIS, one deduces that $s = 0$ and Eqn.~(\ref{Tmms-T}) simply becomes
\begin{equation}\label{eq E-M}
    \mathcal{T}_{s} (\psi, g ,  f) = T (\psi_m , g_m)=   T ( \psi, g ) .
\end{equation}

In the context of large-scale astrophysics, where the energy-momentum tensor in the form of a perfect fluid is more desirable, the Einstein equation is usually written as
    \begin{equation}\label{gr field equ}
		G(g) =  8 \pi \left( T_b(\psi_b , g) + T_d(\psi_d , g) \right) ,
	\end{equation}
wherein the indices $b$ and $d$ stand for baryonic and dark matter fields, respectively. However, comparing Eqns.~(\ref{E-M in Riem}) and~(\ref{gr field equ}), in the framework of MMS, can lead one to conclude   
\begin{equation}\label{Td and H1}
 T_d (\psi_d, g) =  \mathcal{H}(g ,f) ,
	\end{equation}
 and
 \begin{equation}
     T_b (\psi_b, g) = T (\psi_m, g_m)=   T (\psi, g)   .
 \end{equation}
In other words, the result of fixing a conformal frame resembles the effect of an unknown matter field, dubbed dark matter.

\section{ nearly Conformally flat approximation in metric measure space}\label{Linearized theory}

GR does not possess many exact solutions, and the perturbation methods are the practical ways to construct approximate solutions close to the exact ones. In principle, there are no restrictions on the orders of approximation in perturbation methods. However, in some cases, a decomposition of metric at the linear level to $g=g_0+h$ seems useful when components of $h$ are small enough, i.e., with ``the clear separation of scales'' (according to~\cite{maggiore2007gravitational}) in some coordinate system. This helps one, for instance, achieve $h$ as a field in spacetime with background metric $g_0$. Moreover, setting $g_0 \equiv \eta$, the Minkowski metric, resembles the weak field limit, literally known as the linearized theory. Nevertheless, writing down metrics in the specific form
\begin{equation}\label{linearized metric}
	g_{\mu \nu}=\eta_{\mu \nu}+h_{\mu \nu} \hspace{5mm} , \hspace{5mm} |h_{\mu \nu}|\ll 1
\end{equation}
in a particular coordinate system is a kind of gauge fixing that breaks the inherent symmetry of GR, viz., diffeomorphism invariance. It should be noted that although the whole group of diffeomorphism is no longer the symmetry group of the linearized theory, a subgroup of it still preserves the symmetry group for~(\ref{linearized metric}) i.e. $h_{\mu\nu}$ and $h'_{\mu\nu}$ represent the same perturbation field if they admit $h_{\mu\nu} - h'_{\mu\nu} = \epsilon \mathcal{L}_X \eta_{\mu\nu}$ for a given vector field $X$ on the background Minkowski spacetime and any arbitrary infinitesimal constant $\epsilon$ (to guarantee the smallness of $h'$ as well). Nonetheless, GWs, interpreted in this gauge, are not an illusion of gauge fixing and will not fade away by diffeomorphisms. It is worth noting that in Eqn.~(\ref{linearized metric}), one reduces the background to SR but saves the effect of GR in $h$. In other words, even though SR does not predict GWs, it can still be used to demonstrate one of the outcomes of GR. From the mathematical point of view, when~(\ref{linearized metric}) is considered, the infinite-dimensional symmetry group of diffeomorphism has been reduced to a finite-dimensional Poincar\'e group.
	
Recalling that MMS has been introduced as a platform for constructing a gravitational theory with the larger symmetry group, it is useful to derive the linearized version of MMS, in which the conformal symmetry is also broken. This is similar in nature to what has been done above for GR and we follow the same way of reasoning used for the linearized theory except in this case, the theory is reduced to linearized GR with a perturbation that preserves the effect of MMS. In other words, we assume
\begin{equation}\label{decom f}
	f = f_0 + \mathfrak{f} \hspace{5mm} , \hspace{5mm} |\mathfrak{f}|\ll 1 
\end{equation}
where consequently
\begin{equation}\label{expan f}
   e^f \approx 1 + \mathfrak{f} ,
   \end{equation}
in which, for convenience, we set $f_0 = 0$. It is worth noting that there is a subgroup of the conformal group which still presents the symmetry in~(\ref{decom f});  $\mathfrak{f}$ and $\hat{\mathfrak{f}}$ represent the same perturbation in weights if
\begin{equation}\label{f-subg}
	\mathfrak{f} - \hat{\mathfrak{f}}= \mathfrak{w}, 
\end{equation}
with $\mathfrak{w}$ being any arbitrary infinitesimal constant to ensure the smallness of $\hat{\mathfrak{f}}$ as well. Further in this section, we show that~(\ref{f-subg}) is consistent with the degrees of freedom of a scalar potential. Therefore, the full linearized conformal transformation in 4D is completed by
\begin{equation}\label{expan etric}
	g_m = e^{-\mathfrak{f}} g \approx \left( 1-\mathfrak{f} \right) \left( \eta + h \right) ,
\end{equation}
when a class of linearized solutions is regarded. We call Eqn.~(\ref{expan etric}) the nearly conformally flat approximation.

The conclusion to be drawn is that when an infinitesimal coordinate transformation $x^{\mu}\rightarrow x'^{\mu}=x^{\mu} + \xi^{\mu}$ where $\xi^{\mu}\ll1$, and conformal transformation $\hat{g}_{\mu \nu}=e^{2 \omega}g_{\mu \nu}$ where $\omega\ll1$ are applied on metric, some straightforward calculations result in
\begin{equation}\label{hprime}
	h_{\mu \nu} - h'_{\mu \nu}= \left(-2 \omega \eta_{\mu \nu} + \mathcal{L}_{\xi} \eta_{\mu\nu}\right), 
\end{equation}
up to the first order. For the sake of simplicity, we split the modified Einstein's tensor into two parts as shown in Eqn.~(\ref{separated G}); the ordinary Einstein tensor follows the common methods by defining $\bar{h}_{\mu \nu} \equiv h_{\mu \nu} -\frac{1}{2} \eta_{\mu \nu}h$ and using the Lorentz gauge $\partial^{\mu}\bar{h}_{\mu \nu}=0$, and the $\mathcal{H}_{\mu \nu}$ part by applying Eqn.~(\ref{linearized metric}) to Eqn.~(\ref{def H}) yields
\begin{equation}\label{linear H}
  \mathcal{H}_{\mu\nu}(g, f)  = - \frac{1}{8 \pi} \left( \partial_{\mu} \partial_{\nu} \mathfrak{f} -\eta_{\mu \nu}\Box \mathfrak{f} \right) + \mathcal{O}(\epsilon^{2}) .
\end{equation}
		
Furthermore, to obtain the quasi-Newtonian limit of the theory, we apply slow motion approximation which is indicated by $\frac{dx^{i}}{d \tau}=\epsilon \frac{dx^{0}}{d \tau}$. Therefore by adopting $\epsilon \equiv \frac{v}{c}\ll 1$, the slow-motion condition for any function $\mathfrak{f}$ can be written as 
\begin{equation}\label{slow f}
	\frac{\partial \mathfrak{f}}{\partial x^{0}} \sim\epsilon \frac{\partial \mathfrak{f}}{\partial x^{i}},
\end{equation}
demonstrating that as a slowly changing quantity, the changes in the gradient of $\mathfrak{f}$ are much greater than its time evolution. Thus one can assume $\nabla_{i} \mathfrak{f} \propto \mathcal{O}(\epsilon)$ that leads to $ \nabla_{0} \mathfrak{f} \propto \mathcal{O}(\epsilon^2)$.
	
By applying these assumptions, Eqn.~(\ref{linear H}) is reduced to
\begin{equation}\label{H00}
  \mathcal{H}_{00}= - \frac{1}{8 \pi} \eta^{pq}\partial_{p}\partial_{q}\mathfrak{f} + \mathcal{O}(\epsilon^2),
\end{equation}
and
\begin{equation}\label{Hkl}
	\mathcal{H}_{kl}= - \frac{1}{8 \pi} \left(\partial_{k}\partial_{l}\mathfrak{f} - \eta_{kl} \eta^{pq}\partial_{p}\partial_{q}\mathfrak{f} \right) + \mathcal{O}(\epsilon^2) .
\end{equation}

Consequently, the components of the weak counterpart of the Einstein tensor in MMS are     
\begin{equation}\label{TG00}
	\mathcal{G}_{00}=-\frac{1}{2}\nabla^{2}\left(\bar{h}_{00}-2\mathfrak{f}\right),
\end{equation}
and
\begin{equation}\label{TGkls}
	\mathcal{G}_{kl}=-\frac{1}{2}\nabla^{2}\left(\bar{h}_{kl}+2 \eta_{kl}\mathfrak{f}\right)+\partial_{k}\partial_{l}\mathfrak{f} .
\end{equation}
up to the first order in $\epsilon$. 

In the presence of matter with the stress-energy tensor $T_{\mu \nu}=(\rho+P)U_{\mu}U_{\nu}+P g_{\mu \nu}$, recalling Eqns.~(\ref{mod G}) and~(\ref{eq E-M}) and adopting slow motion condition, i.e., $|v|\ll 1$, one can write  $|T_{00}| \gg |T_{0i}|$ and $|T_{00}|\gg |T_{ij}|$~\cite{mtw}. The leading order terms of the weak field equation in MMS theory then is 
\begin{equation}\label{weak EoM}
		-\frac{1}{2}\nabla^{2}\left(\bar{h}_{00}-2\mathfrak{f}\right)=8 \pi \rho.
\end{equation}

 In comparison to the equation of motion of the Newtonian limit (gravity) $\nabla^{2}\Phi=4 \pi \rho$, where $\Phi$ refers to the gravitational potential of the matter distribution, one can write $\bar{h}_{00}=2\mathfrak{f}-4 \Phi$. This simply leads to 

\begin{equation}\label{h00}
	h_{00} =\mathfrak{f}-2 \Phi.	
\end{equation}
Accordingly, some straightforward calculations, using $\bar{h}_{\mu \nu} \equiv h_{\mu \nu} -\frac{1}{2} \eta_{\mu \nu}h$ result in $h_{kl} =(\mathfrak{f}-2 \Phi) \delta_{k l}$. The Eqn.~(\ref{h00}) is consistent with the results presented in~\cite{rahmanpour2016metric}. Finally, the weak metric corresponding to MMS theory is obtained as
\begin{equation}\label{weak metric}
	{ds_{m}}^2=-(1+2 \Phi-\mathfrak{f}) dt^2+(1-2 \Phi+\mathfrak{f})\left(dx^2+dy^2+dz^2\right).   	
\end{equation}

\section{the vacuum solution}\label{Spherical}

In this section, we find a vacuum solution of gravitational theory in MMS, which corresponds to the spherically symmetric vacuum solution of the 4D spacetime of GR. By this, we mean there is a canonical metric $ g_{m}$ in MMS with 
\begin{equation}\label{vacuum G mms}
	 \mathcal{G}_{s}(g_{m},0)=0.
\end{equation}
 Using two essential features of constructing the conformally invariant tensors in MMS, namely, $\mathcal{I}_{s}(g,0)=I(g)$, and, $\mathcal{I}_{s}(\hat{g},\hat{f})= \exp(s\omega) \mathcal{I}_{s}(g,f)$, one has
\begin{equation}\label{general equiv G}
	\mathcal{G}_{s}(g_{m},0)=  G(g_m),
\end{equation}
which by applying Eqns.~(\ref{vacuum G mms}), results in 
\begin{equation}\label{vacuum G}
	G(g_m)=0,
\end{equation}
where $ g_{m}$ is the Schwarzschild metric
\begin{equation}\label{isotropic}
	{ds_{m}}^2 = - \frac{\left( 1 - \frac{C_0}{4 r} \right)^2}{\left( 1 + \frac{C_0}{4r} \right)^2} dt^2 + \left( 1 + \frac{C_0}{4r} \right)^{4} \left({dr}^2 + {r}^2 d\Omega^2\right),
\end{equation}
in the isotropic coordinates. In this equation $C_{0}$ is a constant and the index $m$ denotes the canonical metric. As mentioned before and regarding the coincidence principle, GR should be retrieved by setting $f=0$. Then according to GR, Eqns.~(\ref{vacuum G}) and~(\ref{isotropic}) together lead to  $C_{0}=2 M$.

On the other hand Eqn.~(\ref{vacuum G mms}), for any other metric which is conformally related to $g_{m}$ via $g=e^{f} g_{m}$ yields
\begin{equation}\label{vac separated G}
	G_{\mu\nu}(g) = 8 \pi \mathcal{H}_{\mu\nu}(g,f).
\end{equation}
 For the sake of simplicity, if we restrict $f$ to be a function of radial coordinate only, the spherical symmetry of the vacuum solution in MMS is retained. Therefore, one can assume $g$ to be of the form 
\begin{equation}\label{ss metric}
	ds^2 = - e^{2\alpha (r)} dt^2 + e^{2\beta (r)}\left( dr^2 + r^2 d\Omega^2\right),
\end{equation}
which is related to $g_{m}$ by
\begin{equation}\label{Gm}
	{ds_{m}}^2 = - e^{2\alpha (r)-f(r)} dt^2 + e^{2\beta (r)-f(r)}\left( dr^2 + r^2 d\Omega^2\right).
\end{equation}
By comparing Eqns.~(\ref{isotropic}) and ~(\ref{Gm}) of this gravitational theory, one can obtain components of the metric
\begin{equation}\label{alpha}
	2\alpha (r)= f+2 \ln \left(\left(1-\frac{2 M}{4r}\right)-\left( 1+\frac{2 M}{4r}\right)\right),
\end{equation}
\begin{equation}\label{beta}
	2\beta (r)= f+4 \ln \left(1+\frac{2 M}{4r}\right).
\end{equation}

 It is worth noting that, while in a conformal frame $(g_{m},0)$,  Eqn.~(\ref{vacuum G}) reminds one of vacuum Einstein's field equation, in another conformal frame $(g,f)$, Eqn.~(\ref{vac separated G}) simulates an illusion of matter in GR. We employ this idea in its weak field limit to explain the FRC of galaxies, as we shall investigate in the next section. For further convenience, we note that the weak field approximation of Eqns.~(\ref{alpha}) and (\ref{beta}) are 
\begin{equation}\label{weak alpha}
	2\alpha (r) =\mathfrak{f}-\frac{2 M}{r}
\end{equation}
and
\begin{equation}\label{weak beta}
	2\beta (r)= \mathfrak{f}+\frac{2 M}{r}.
\end{equation}

\section{ resolution for FRC problem}\label{RotCur}

Briefly, the FRC problem is a disagreement between the prediction of Newtonian gravity and observational data; although the former predicts the rates of tangential velocities of rotating stars to be decreasing by increasing radius in the outer regions of the galactic bulge, the latter shows nearly constant values~(see for example \cite{bertone2018history} and references therein). In other words, considering the amount of visible mass in galaxies only, the stars should rotate slower than what is observed~\cite{oort1932force}. Therefore, a kind of unknown mass distribution, namely dark matter, may be responsible for it. However, there are several conceivable scenarios to explain this discrepancy. In this section, we use the idea discussed in the previous sections that a vacuum solution of gravitational theory in MMS, when restricted to the context of GR (in a conformal symmetry broken frame), can simulate an illusion of a matter field, dubbed as $\mathcal{H}$ in this manuscript, to justify FRC. For an appropriate density function, $\mathcal{H}$ can play the role of dark matter and explain FRC. It is customary to use the spherically symmetric metric in order to model FRC in the simplest form. To do this, the Newtonian gravitational potential $\Phi(r)$, appeared in metric
\begin{equation}\label{sph sym met}
	{ds}^2 = - e^{2\Phi (r)} dt^2 + e^{2\gamma (r)} \left(dr^2 + r^2 d\Omega^2\right),
\end{equation}
is tuned to explain FRC and consequently is supposed to be produced by dark matter distribution. Instead, here we seek a density function $f(r)$ that conformally transforms this metric into the vacuum solution of GR, recalling that these two apparently different solutions are the same in the framework of MMS. As mentioned before, to respect the spherical symmetric nature of the vacuum solution in MMS, we consider the density function to be a function of radial coordinate only. This is consistent with the observed more or less spherically symmetric nature of FRC. 

Moreover, since FRC can be explained in the Newtonian regime, we use the benefit of the weak field approximation Eqns.~(\ref{weak alpha}) and ~(\ref{weak beta}) in consistency with~(\ref{weak metric}), and find 
\begin{equation}\label{equality}
	\frac{-2 M}{r}= 2\Phi(r)-\mathfrak{f}(r)
\end{equation}
and
\begin{equation}\label{equality of beta}
	\frac{2 M}{r}= 2\gamma(r)-\mathfrak{f}(r),
\end{equation}
 wherein $\alpha(r)$ and $\beta(r)$ are replaced by $\Phi(r)$ and $\gamma(r)$, respectively. Using 
\begin{equation}\label{v_t}	
	{v_{t}}^{2}= \frac{r \Phi^{\prime}(r)}{1+r \gamma^{\prime}(r)},
\end{equation}
 derived in appendix~\ref{tan vel}, and where $v_{t}(r)$ is the tangential velocity of stars rotating around the galactic center coming from observations, Eqn.~(\ref{equality}) eventually leads to the general form of the density function
\begin{equation}\label{f(r)}
\mathfrak{f}(r)=\int_{r_0}^{r} dr' \frac{2}{1-{v_{t}}^{2}(r')} \left[\frac{{v_{t}}^{2}(r')}{r'} - (1+{v_{t}}^{2}(r'))\frac{M}{r'^{2}}\right].
\end{equation}
in which $r_{0}$ is an arbitrary radius used  to denote the origin of the density function. The choice of $r_{0}$, which globally shifts $\mathfrak{f}(r)$ by a constant, corresponds to a global rescaling and is a matter of convenience. According to Eqn.~(\ref{f(r)}), for any tangential velocity profile, one can obtain the corresponding density function that accounts for the flattening of the rotation curve. For the plausible assumption $v_{t}\ll 1$ which is consistent with the slow-motion condition, Eqn.~(\ref{f(r)}) reduces to 
\begin{equation}\label{rduced f(r)}
\mathfrak{f}(r)= \frac{2M}{r} - \frac{2M}{r_{0}} + 2 \int_{r_0}^{r} dr'  \frac{{v_{t}}^{2}(r')}{r'}.
\end{equation}

In the following subsections, we try to find the form of the corresponding density functions, beginning from a naive constant tangential velocity model and proceeding to two well-known PSS and NFW profiles of velocity and mass density, respectively. 
\subsection{constant velocity profile}\label{const profile}

 By way of explanation, we start with the simplest case in which the tangential velocity does not depend on coordinates. Then Eqn.~(\ref{rduced f(r)}) simply yields
\begin{equation}\label{f const}
\mathfrak{f}_{c}(r) =\frac{2 M}{r} - \frac{2 M}{r_{0}} +2 {v_{t}}^{2} \ln (\frac{r}{r_{0}}). 
\end{equation}
 For this spherically symmetric toy model, the entire baryonic mass of a galaxy, including the stellar and gas, is considered to be clustered in a sphere with the radius $R_{b}$ and mass $M_{b}$, which are called the radius and mass of concentrated baryonic matter, respectively. Additionally, we expect the flattening of the galactic rotation curve to occur outside of this radius. Supposing the start point of FRC is where $v_{t}(r)=(\frac{M_{b}}{R_{b}})^{1/2}$, The density function~(\ref{rduced f(r)}) can be reduced to
\begin{equation}\label{f for v const1}
	\mathfrak{f}_{c}(r)=\frac{2 M_{b}}{r}+ \frac{2 M_{b}}{R_b} \ln \left(\frac{r}{R_b}\right).
\end{equation}
where the constant term $-\frac{2 M_{b}}{R_{b}}$, responsible for a global rescaling, is ignored. It is worth noting that for galaxies whose most of the mass is accumulated in their spherical bulge, the above parameters might be the mass and the radius of their central bulges. 
In other words, having the parameters $M_{b}$ and $R_{b}$ for a galaxy, Eqn.~(\ref{f for v const1}) can describe FRC as a part of the geometry that can eliminate the need for dark matter. 

\subsection{Persic--Salucci--Stel (PSS) velocity profile}\label{pss profile}
The Persic--Salucci--Stel (PSS) is a velocity profile to describe FRC proposed by Persic et al.~\cite{persic1996universal}. They claim their universal profile strongly depends on the luminosity of the galaxies, which shows a slight discrepancy for the bright galaxies while in good accordance with low luminous ones. The velocity profile is
\begin{equation}\label{pss vel1}
	{v_{PSS}}(r)=\left(\frac{{v_{0}}^2 r^2}{r^2+{R^2}}\right)^{1/2},
\end{equation}
with two fitting parameters, $v_{0}$ and $R$, viz, the 'terminal velocity' and 'the velocity core radius',  respectively~\cite{barranco2015dark,persic1996universal}. The density function related to the PSS velocity profile, 
utilizing Eqn.~(\ref{rduced f(r)}), is 
\begin{equation}\label{f pss}
\mathfrak{f}_{PSS}(r)=\frac{2 M}{r} -\frac{2 M}{r_{0}} + v_{0}^{2} \ln \left(\frac{r^{2}+R^{2}}{r_{0}^{2}+R^{2}}\right).
\end{equation}
To lower the number of parameters, we set $r_{0}=R$ then remove the constant terms, similar to what has been done in subsection ~\ref{const profile}, and find 
\begin{equation}\label{f pss2}
\mathfrak{f}_{PSS}(r)=\frac{2 M}{r} + v_{0}^{2} \ln \left(1+\frac{r^{2}}{ R^{2}}\right).
\end{equation}

\subsection{Navarro--Frenk--white (NFW) density profile}\label{nfw profile}

One of the most commonly used mass density profiles to describe dark matter halo is the NFW profile~\cite{navarro1996structure}. This universal profile is obtained from N-body simulations and describes the dark halo density as a function of radial coordinate, which varies from the inner to the outer parts of the galaxy~\cite{lokas2001properties}. The profile is 
\begin{equation}\label{nfw}
	\rho_{NFW}(r)=\frac{\sigma}{r(r+R_s)^2},
\end{equation}
with $\sigma \equiv \rho_{cri} \delta_{c} {R_{s}}^3$, where $\rho_{cri}$ and $R_{s}$ are called the critical density of the universe and the characteristic radius, respectively and $\delta_{c}$ is a dimensionless parameter. Solving the corresponding Poisson's equation and substituting the gravitational potential to Eqn.~(\ref{rduced f(r)}), the density function according to this profile becomes
\begin{equation}\label{f nfw}
    \mathfrak{f}_{NFW}=\frac{2 M}{r} + \frac{8 \pi \sigma}{r} \ln \left(1+\frac{r}{R_{s}}\right),
\end{equation}
following the similar procedure for Eqns.~(\ref{f for v const1}) and (\ref{f pss2}).

\section{Conclusion and Remarks}\label{ConRem}
In this manuscript, by gravitational theory in MMS we mean a gravitational theory that, in addition to diffeomorphism invariance, respects the conformal invariance. To compensate for the side-effect of Weyl rescaling of metric, the volume element is subject to change consistently by the introduction of a density function. These joint transformations, simply called conformal transformations, enable us to promote the pseudo-Riemannian invariants of GR to conformal invariant ones.

Although the density function plays the role of a gauge and appears as a degree of freedom, one is not restricted to assuming it as a pure gauge. For instance, the platform of MMS may be a fruitful framework for theories that can fix the density function for other reasons; non-local gravity theories could be the case. It is worth noting that in MMS, the matter field is no longer forced to be traceless.

Although there are considerable debates in the literature about the perforation of either larger or smaller symmetry groups of a theory to be, the aspect of conformal invariance, when added to a gravitational theory like GR, seems to widen the strength and the realm of the theory as well. For example, it may equip the theory in a way that becomes more suitable for quantization requirements such as renormalization. Also, it is deemed to be more suitable as a theory to unify physics from small to large scales. 

In this work, we have concentrated on the vacuum solution of MMS. Not surprisingly, this solution when regarded in different conformal frames is generally equivalent to different solutions of GR.  Following this idea, we suggest how an appropriate conformal frame can resemble the illusion of a dark matter field and deviation of the rotation curve of galaxies from what is expected in the Newtonian regime. In other words, this phenomenon is a (conformally) frame-dependent observation. 
This encourages one to seek a similar explanation for dark energy in further studies. 

\appendix
\begin{appendices}

\section{}\label{tan vel}
	In this appendix, following~\cite{chandrasekhar1998mathematical,matos2002geometric}, we try to derive the  velocity of a test particle rotating in a circular orbit in the spherically symmetric and static spacetime represented by an isotropic line element
	\begin{equation}\label{a1}
		ds^2=-e^{2 \Phi(r)}dt^2+e^{2 \gamma(r)}\left(dr^2+r^2d\Omega^2\right).
	\end{equation}
 We aim to find the velocity and consequently the corresponding Newtonian potential field in the context of GR. Hence the Lagrangian 
	\begin{equation}\label{a2}
		\mathcal{L}=\frac{1}{2}g_{\mu \nu} \frac{\mathrm{d} x^{\mu}}{\mathrm{d} \tau} \frac{\mathrm{d} x^{\nu}}{\mathrm{d} \tau},
	\end{equation}
	 reads as
	\begin{equation}\label{a3}
		\mathcal{L}=\frac{1}{2}\left[-e^{2 \Phi(r)} \dot{t}^2+e^{2 \gamma(r)} \left(\dot{r}^2+r^2 \left(\dot{\theta}^2+\sin^{2}\theta \dot{\phi}^2 \right)\right)\right],
	\end{equation}
	where the dot stands for the derivative with respect to $\tau$, which is the proper time of a particle moving along the geodesic. By defining $E\equiv e^{2 \Phi(r)} \dot{t}$, $L_{\theta} \equiv  e^{\gamma(r)} r^2 \dot{\theta}$ and $L_{\phi} \equiv e^{\gamma(r)} r^2 \sin^2 \theta \dot{\phi}$, and setting $2 \mathcal{L}=-1$ for a timelike geodesic, one can write the radial equation of motion
	\begin{equation}\label{a4}
		\dot{r}^{2}+V(r)=0,
	\end{equation}
	with the potential $V(r)$ 
	\begin{equation}\label{a5}
		V(r)=-e^{-2 \gamma(r)}\left( e^{-2 \Phi(r)}E^{2}- r^{-2} e^{-2 \gamma(r)}L^{2} -1\right),
	\end{equation} 
	wherein $E$ and $L^{2}={L_{\theta}}^{2}+\left(\frac{ L_{\phi}}{\sin \theta}\right)^{2}$ are considered as the total energy and angular momentum, respectively.
	
	Employing the conditions of stability for a test particle rotating in a circular orbit, the total energy and angular momentum in terms of the coefficients of metric are
	\begin{equation}\label{a6}
		\begin{aligned}
		&{E^2}=\frac{e^{2 \Phi(r)}\left(1+r \gamma^{\prime}(r)\right)}{1-r \Phi^{\prime}(r)+r \gamma^{\prime}(r)},\\
        &L^{2}=\frac{e^{2 \gamma(r)} r^{3} \Phi^{\prime}(r)}{1-r \Phi^{\prime}(r)+r \gamma^{\prime}(r)},
		\end{aligned}
	\end{equation}
	with prime denotes the derivative with respect to $r$. To find the four-velocity of a particle moving along its time-like worldline, we set $d\tau ^{2}=-ds^2$ in Eqn.~(\ref{a1})
	\begin{equation}\label{a7}
		d\tau^2=e^{2 \Phi(r)}dt^2 \left[ 1-e^{2 \gamma(r)-2 \Phi(r)}\left(\left(\frac{dr}{dt}\right)^{2} + r^2 \left(\frac{d\Omega}{dt}\right)^{2}\right)\right],
	\end{equation} 
	wherein the tangential velocity $ {v_{t}}^{2}=e^{2 \gamma(r)-2 \Phi(r)} r^2 \left(\frac{d\Omega}{dt}\right)^{2}$ with a straightforward calculation, can be written as
	\begin{equation}\label{a8}	
		{v_{t}}^{2}= r^{-2}e^{2 \Phi(r)-2 \gamma(r)} L^{2} E^{-2}= \frac{r \Phi^{\prime}(r)}{1+r \gamma^{\prime}(r)},
	\end{equation}
in terms of the coefficients of metric.

\end{appendices}
\bibliography{ref}

\end{document}